\documentclass[twocolumn,superscriptaddress,prd,eqsecnum]{revtex4}
\usepackage{amsmath,amssymb}

\usepackage{graphicx}

\begin{document}
\title{RE-INTERPRETATION OF SPONTANEOUS SYMMETRY BREAKING
IN QUANTUM FIELD THEORY 
AND GOLDSTONE THEOREM }

\author{T. \textsc{Fujita}} \email{fffujita@phys.cst.nihon-u.ac.jp} 
\author{M. \textsc{Hiramoto}}  
\author{T. \textsc{Homma}} 
\author{M. \textsc{Matsumoto}}
\author{H. \textsc{Takahashi}} \email{htaka@phys.ge.cst.nihon-u.ac.jp}

\affiliation{
Department of Physics, Faculty of Science and Technology 
Nihon University, Tokyo, Japan
}

\date{\today}

\begin{abstract}
We present a new picture of global symmetry breaking in quantum field theory
and propose a novel realization of symmetry breaking phenomena in terms of 
the conserved charge associated with its symmetry. 
In particular, the fermion condensate of the vacuum state is examined 
when the spontaneous chiral symmetry breaking takes place. 
It is shown that the fermion condensate of the vacuum  
vanishes if the system is solved exactly, and therefore 
we cannot make use of the Goldstone theorem. 
As a perfect example, we present the Bethe ansatz vacuum 
of the Thirring model which shows the spontaneous chiral 
symmetry breaking with no fermion condensate. 
\end{abstract}

\maketitle

\section{Introduction}
The spontaneous symmetry breaking has been extensively studied in particle physics 
as well as in condensed matter physics. The physics of the symmetry breaking is related 
to many interesting phenomena such as the phase transition of ferromagnetism, 
dynamical mass generation, Higgs mechanism and so on. 

In fermion field theory, the chiral symmetry breaking has played 
an important role for understanding the vacuum structure, 
and indeed it has been discussed repeatedly by many people \cite{q1,q2,q7,q07}. 
The importance of the chiral symmetry and its breaking 
is related to the non-perturbative aspect of the vacuum of the field theory. 
The property of the spontaneous chiral symmetry breaking cannot be discussed 
in the perturbation theory, but it should be evaluated in an exact or 
non-perturbative fashion.  

The chiral symmetry broken vacuum is first obtained by Nambu and Jona-Lasinio \cite{q7} 
who solved the current current interaction model (NJL model) 
in terms of the Bogoliubov transformation technique, 
and this is indeed a non-perturbative method.  However, it is, at the same time, 
clear that the Bogoliubov vacuum is not exact, that is, it is not the eigenstate 
of the  Hamiltonian even though it may give a good description of 
the spectrum of the field theory model. In this sense, we have to carefully examine 
the calculated results of the NJL model concerning the physics related to the spontaneous 
chiral symmetry breaking. 

In two dimensions, the situation becomes quite different for the symmetry breaking 
phenomena from the four dimensions. There is the Coleman theorem \cite{q3}. 
Since a massless scalar field is not well defined in two dimensions because 
of the infrared singularity of the massless scalar field, the Nambu-Goldstone (NG) boson 
cannot exist. Due to the Goldstone theorem, Coleman concluded that the spontaneous 
symmetry breaking should not occur in two dimensional field theory. 
Here, it is important to note that the Coleman theorem is meaningful only if 
the Goldstone theorem holds true.

However, recent careful studies clarify that there appears no massless boson after 
the spontaneous chiral symmetry breaking in fermion field theory 
models \cite{q9,q91,qq1,p5,hist,nova}. 
There, it is shown that the Goldstone theorem only proves 
that there must be a free massless fermion and anti-fermion pair, instead 
of the NG boson if one starts from the finite fermion condensate value. 

In addition, the spontaneous chiral symmetry breaking takes place in the Thirring 
model as well as in QCD$_2$ which are in fact two dimensional field theory models. 
Further, the calculations of bosons after the spontaneous chiral symmetry breaking 
in the fermion field theory models of the NJL, Thirring and QCD$_2$ predict always 
a finite boson mass or no bosonic excitation (only a gap in the excitation spectrum) 
as long as one starts from the proper symmetry broken vacuum state. For the boson mass 
calculation after the symmetry breaking, one should take the true vacuum state which 
breaks the symmetry. It should be noted that, if one takes the perturbative 
vacuum  \cite{nova} for the boson mass evaluation, then one obtains unphysical spectrum. 

In this paper, we reformulate the global symmetry breaking phenomena of quantum field 
theory in terms of the eigenvalue problem of the conserved  charge associated 
with the symmetry. In particular, it is shown that the fermion condensate of 
the fermion field theory models must vanish for the exact vacuum which breaks the chiral 
symmetry spontaneously. Here, we show a good example of the spontaneous chiral symmetry 
breaking which satisfies all the above conditions with zero fermion condensate. 
This is the Bethe ansatz solution of the massless Thirring model, and the exact 
eigenstate of the vacuum is described analytically \cite{p5,non}. 
This vacuum shows that the chiral symmetry is broken, but 
there is no fermion condensate. The spectrum of the Thirring model is also known, and 
the Bethe ansatz solutions show that the excitation spectrum of the Thirring model 
has a finite gap, but there is no boson. 

Further, if we solve the Thirring model in terms of the Bogoliubov transformation 
method, then we find that the Bogoliubov vacuum has a finite condensate \cite{q9,q91,q10}. 
Clearly, the finite condensate arises from the approximation made in the Bogoliubov 
vacuum state which is not the eigenstate of the Hamiltonian. 
Therefore, it is important to understand that the symmetry breaking has nothing to do with 
the fermion condensate since the condensate vanishes in the true vacuum state. 
In this respect, one sees that the fermion condensate acts as the indicator 
of the approximation rather than the index of the symmetry breaking.  

This paper is organized as follows. 
In the next section, we critically review the shortcomings of the common understanding 
of the symmetry breaking physics as well as Goldstone theorem. In section \ref{new_sbp}, 
we reformulate the symmetry breaking physics in terms of the conserved charge. 
Further, we explain the Thirring model as an example of the chiral symmetry breaking. 
Section \ref{sec_chb_cond} describes the condensate in the chiral symmetry breaking 
phenomena. We show that the condensate value vanishes in the exact vacuum state in 
the Thirring model. Also we reexamine the validity of the Coleman theorem. 
In section \ref{condensate_gauge}, we discuss 
the condensate value of the two dimensional 
gauge theories. Section \ref{conclusion} summarizes what we have learned about 
the spontaneous symmetry breaking physics in this paper.

\section{Old picture of symmetry breaking }
\subsection{Charge of symmetry}
In this section, we review the symmetry breaking phenomena in quantum field theory
which has been discussed repeatedly until now. Here, the basic notations concerning 
the spontaneous symmetry breaking physics are introduced. 

Suppose that $D$-dimensional system is invariant under the global symmetry group ${\cal G}$.
In this case, there is a conserved current $j_\mu^a$ ($a = 1, 2, \cdots, \textrm{dim} {\cal G}$) which satisfies
\begin{equation}
\partial^\mu j_\mu^a = 0 .
\end{equation}
The Hamiltonian $H$ of the system is also invariant under the symmetry ${\cal G}$ whose 
unitary transformation is given by
\begin{equation}
U_a(\alpha) = e^{i\alpha \hat {\cal Q}_a} ,
\end{equation}
where
\begin{equation}
\hat {\cal Q}_a = \int \!d^D x \,j_0^a (x) .
\end{equation}
Here, we can write
\begin{equation}
U_a(\alpha)H U_a(\alpha)^{-1} =H .
\end{equation}
Therefore, the charge $\hat {\cal Q}_a$ of the symmetry ${\cal G}$ commutes 
with the Hamiltonian $H$,  
\begin{equation}
\hat {\cal Q}_a H = H \hat {\cal Q}_a .
\label{old_hq_comm}
\end{equation}
Now, the vacuum state can break the symmetry, and we define the symmetry unbroken vacuum 
$|0\rangle $ and  symmetry broken vacuum $|\Omega \rangle $, respectively, which 
satisfy the following equations, 
\begin{subequations}
\begin{equation}
 U_a(\alpha) |0\rangle =|0\rangle ,
\end{equation}
\begin{equation}
 U_a(\alpha) |\Omega \rangle \neq |\Omega \rangle .
\end{equation}
\end{subequations}
These equations can be written in terms of the charge operator $\hat {\cal Q}_a$ as
\begin{subequations}
\label{sym_charge_judg}
\begin{equation}
\hat {\cal Q}_a |0\rangle =0  ,
\end{equation}
\begin{equation}
\hat {\cal Q}_a |\Omega \rangle \not= 0 . 
\end{equation}
\end{subequations}
Now, suppose the $|\Omega \rangle $ has the lowest eigenvalue $E_\Omega$.
From eq.(\ref{old_hq_comm}), we have
\begin{equation}
H \hat {\cal Q}_a |\Omega \rangle = E_\Omega \hat {\cal Q}_a|\Omega \rangle .
\end{equation}
Therefore, the broken vacuum is degenerate.
Further, we see that
\begin{align*}
\langle Q_a | Q_a  \rangle = \langle \Omega| \hat {\cal Q}_a^\dagger \hat {\cal Q}_a|\Omega \rangle 
 = \int\!d^Dx \,\langle \Omega| j_0^a(\bm{x}, t) \hat {\cal Q}_a |\Omega \rangle.
\end{align*}
In terms of the translation property, we have
\begin{equation}
\langle Q_a | Q_a  \rangle = \int\!d^Dx \,\langle \Omega| j_0^a(\bm{0}, t) \hat {\cal Q}_a |\Omega \rangle.
\label{old_vac_deg}
\end{equation}
However, the right hand side of eq.(\ref{old_vac_deg}) should not be finite 
when symmetry is broken. 
Therefore, we may conclude that the charge of broken symmetry is not 
well-defined \cite{fabri_pica,itz_zub}.

On the other hand, we also have the relation
\begin{equation}
\langle 0 | \hat {\cal Q}_a |0 \rangle = \int\!d^Dx \,\langle 0| j_0^a(\bm{0}, t)
 |0 \rangle. 
\label{old_vac_unbroken}
\end{equation}
In this case, the right hand side of eq.(\ref{old_vac_unbroken}) should not be finite 
even though \textit{symmetry is unbroken}. 
Now, it is clear that the illness of eq.(\ref{old_vac_deg}) 
and eq.(\ref{old_vac_unbroken}) has nothing to do with the property of 
the charge operator. 
Further, the operator of the system should have the meaning only with the states. 
Therefore, the commutation relation (\ref{old_hq_comm}) between Hamiltonian $\hat H$ and 
broken charge $\hat Q_a$ does not change before and after the symmetry breaking. 
In this sense, the property of the charge operator does not change even after the symmetry 
breaking.  In this respect, the statement that the charge operator is ill-defined 
does not make sense.

\newpage
\subsection{Goldstone theorem}
The Goldstone theorem claims that, when a continuous symmetry is spontaneously broken, 
a massless scalar field should emerge. This massless boson is called Nambu-Goldstone(NG) 
boson \cite{q1,q2,q7}. The outline of the proof of the theorem is the following. 
When the symmetry is broken, the charge $\hat {\cal Q}$ of the symmetry should satisfy 
\[ \hat {\cal Q} | \Omega  \rangle \neq 0 .\]
This should be equivalent to the existence of the relation
\begin{subequations}
\begin{equation}
 \delta \phi(x) = i\left[\hat {\cal Q}, \,\phi(x) \right] = i\int\!d^4y \, 
\bigl[j_0(y), \,\phi(x) \bigr],
\end{equation}
where $\phi(y)$ is \textit{assumed} to be a scalar field operator and
\begin{equation}
\Phi_0 \equiv \langle \Omega| \delta \phi(x) |\Omega \rangle \neq 0.
\label{old_scalar_ev}
\end{equation}
\end{subequations}
Suppose that $\phi(x)$ can be written in terms of the invariant delta function 
$\Delta(x-y)$ of the scalar field 
 (K\"allen-Lehmann representation) \cite{itz_zub,weinberg}
\begin{align}
  \Delta(x-y) &\equiv\left< \Omega\right \lvert[\phi(x), \,\phi(y)] \left\lvert \Omega \right> \nonumber\\
   & =\int_0^\infty \!d\sigma^2 \rho(\sigma^2) i \Delta^{(0)}(x-y; \sigma^2)
\label{old_ngb_spec_rep}
\end{align}
where $\rho(\sigma^2)$ is the spectral function and
\begin{equation}
i \Delta^{(0)}(x-y; \,m^2)\equiv \left<\Omega \right\lvert[\phi_f(x), \,\phi_f(y)]
   \left\lvert \Omega \right>
\end{equation}
is the invariant delta function  of the \textit{free} scalar field $\phi_f(x)$ 
with mass $m$.

Now, we consider the vacuum expectation value 
$\left<\Omega \right\lvert[j_\mu(x), \,\phi(y)] \left\lvert \Omega \right>$.
It can be easily verified in terms of the K\"allen-Lehmann representation of scalar 
field eq.(\ref{old_ngb_spec_rep}) that \cite{q2,weinberg}
\begin{subequations}
\begin{equation}
\left<\Omega \right\lvert \bigl[j_\mu(x), \,\phi(y) \bigr] \left\lvert \Omega \right>
  =\int_0^\infty \!d\sigma^2 \rho(\sigma^2) i \partial_\mu 
   \Delta^{(0)}(x-y; \sigma^2).
\end{equation}
Further,
\begin{equation}
\left<\Omega \right\lvert \textrm{T} j_\mu(x)\,\phi(y) \left\lvert \Omega \right>
  =\int_0^\infty \!d\sigma^2 \rho(\sigma^2) i \partial_\mu  \Delta^{(0)}_F(x-y; \sigma^2),
\label{old_gold_spct_j}
\end{equation}
where $\Delta^{(0)}_F(x-y; \, m^2)$ is the Feynman propagator of a massive scalar boson 
which is given by
\begin{equation}
\Delta^{(0)}_F(x-y; \, m^2) = \int \dfrac{d^4k}{i(2\pi)^4} \dfrac{e^{-ik \cdot (x-y)}}{m^2-k^2-i \epsilon}.
\label{old_ngb_feynmann}
\end{equation}
\end{subequations}
On the other hand, we have
\begin{align*}
\langle \Omega| \delta \phi(y) |\Omega \rangle  
 &= \langle \Omega|i\left[\hat {\cal Q}, \,\phi(y) \right] |\Omega \rangle \\
 &= \int \! d^4x \, i \partial_\mu \left<\Omega \right\lvert \textrm{T} j_\mu(x)\,\phi(y) \left\lvert \Omega \right> .
\end{align*}
Finally, eq.(\ref{old_gold_spct_j}) becomes
\begin{equation}
\Phi_0 = - \lim_{k \rightarrow 0} \int_0^\infty \!d\sigma^2 \dfrac{k^2\rho(\sigma^2)}{\sigma^2-k^2-i \epsilon} .
\label{old_spec_lim}
\end{equation}
Since we have assumed $\Phi_0 \neq 0$, the spectral function can be written by
\begin{equation}
\rho(\sigma^2) = \Phi_0 \delta(\sigma^2) + \rho'(\sigma^2),
\label{old_ngb_spect_fun}
\end{equation}
where $\rho'(\sigma^2)$ is some function which vanishes in the limit 
eq.(\ref{old_spec_lim}). 
In general, the spectral function can be written separable in terms of the one particle 
state  whose mass is $m$ for the scalar field as
\begin{equation}
\rho(\sigma^2) = Z \delta(\sigma^2-m^2) + \rho'(\sigma^2),
\end{equation}
where $Z$ is a constant with $0 \le Z < 1$. 

Correspondingly, if the symmetry is broken or the vacuum expectation value of 
the scalar field has finite value,
\[
 \langle \Omega| \delta \phi(x) |\Omega \rangle \neq 0 ,
\]
then one expects  a massless boson.

\subsection{Coleman theorem}
In two dimensions, the Feynman propagator (\ref{old_ngb_feynmann}) becomes
\begin{equation}
\Delta^{(0)}_F(x; \, m^2) = \int \dfrac{dk_1}{4\pi i} \dfrac{e^{-ik \cdot x}}
{\sqrt{k_1^2+m^2}}.
\end{equation}
Therefore, the Feynman propagator of a massless scalar field has an infrared singularity. 
In this case, according to Coleman, 
the vacuum expectation value of the scalar field operator $\delta \phi$ becomes
\begin{equation}
\Phi_0 = 0
\end{equation}
owing to the Wightman axiom. It means that the spectral function (\ref{old_ngb_spect_fun}) 
has no massless distribution. Therefore, we can conclude with help of Goldstone theorem 
that the symmetry breaking does not occur in two dimensions because of absence of 
a massless boson.

However, it is well-known that there is a counter example of Coleman theorem. 
The chiral invariant Gross-Neveu model, which is the two dimensional version of 
the NJL model,  has the chiral symmetry broken vacuum and the massless fermion 
acquires an effective mass  \cite{gross_neveu}. 
One attempt to resolve this difficulty is that it may be 
the Berezinski-Kosterlitz-Thouless phenomena \cite{witten:npb145}. 
However, there is another difficulty of existence of the massless boson in massless 
QCD$_2$ in which 't Hooft predicted a massless boson in the $1/N$ expansion technique. 
This difficulty should be partially caused by the formulation of the auxiliary field and 
the intrinsic problem of the $1/N$ expansion \cite{fuji_htaka_large} and therefore, 
there should exist no massless boson in QCD$_2$ if calculated properly \cite{qq1}. 
Further, recent investigations of the massless Thirring model show that the chiral 
symmetry of the Thirring model is indeed broken \cite{faber_ivanov_epjc,p5} spontaneously 
without any boson. 

This contradiction will be resolved when we examine the Goldstone theorem and 
solve the intrinsic problem of the theorem.

\subsection{Intrinsic problem of Goldstone theorem}
As seen in the previous subsection, the Goldstone theorem has been considered  
to be valid in the models which have a broken symmetry. 
It seems that the Nambu-Goldstone boson always appears after the spontaneously symmetry 
breaking. 

However, it should become a completely different story when we consider 
the fermion field theory models. 
In fermion field theory models, i.e. the NJL model and Thirring model, for examples, 
there is no bosonic degree of freedom in the Lagrangian density. Therefore, the boson 
state must be constructed as a bound state of fermions.

Why can the Goldstone theorem predict a massless boson state without considering 
the bound state problem ? This is simply because \textit{it is a priori assumption 
in the Goldstone theorem}.  That is, the existence of the scalar field operator 
$\phi$ (here, we called it NG field) as well as the spectral property of the boson field, 
which is represented in terms of eq.(\ref{old_ngb_spec_rep}), 
\textit{is assumed without considering any physical verification}.
However, for the case of NJL model and Thirring model, we cannot accept such a assumption 
since there is no boson field. 
Therefore, we must prove the theorem without eq.(\ref{old_ngb_spec_rep}) 
for the case of fermion field theory models. 

For the NJL model, Nambu and Jona-Lasinio considered the bound state problem in terms of 
the Bethe-Salpeter equation \cite{q7} and claimed that there is a NG boson after 
the spontaneous symmetry breaking.  Unfortunately, however, their calculation of 
the boson spectrum was based on the perturbative vacuum even though the real vacuum 
is not perturbative  one after the symmetry breaking occurred. Therefore, their proof 
of the existence of a massless boson cannot be justified \cite{hist}. 

On the contrary, careful calculations in terms of the Bogoliubov method, 
which is the same as Nambu and Jona-Lasinio's formulation,  
show that the boson state becomes always massive if one employs the symmetry broken 
vacuum state or proper Bogoliubov vacuum state. Further, the Thirring model of 
two-dimensional fermion model has also a massive boson after the spontaneous symmetry 
breaking.

It should be noted that, even though  the Bogoliubov transformation method 
predict that there exists a massive boson in the NJL and Thirring models 
after the chiral symmetry breaking, the existence of the massive boson itself is 
not a confirmed statement since the Bogoliubov transformation method is only 
an approximate scheme. Naively, fermions move with the speed of light and 
the interaction of the Thirring model is a delta function type. 
Further, there is no gauge field. That is, the interaction between the fermions is 
effectively the elastic scattering. Therefore, it should be quite difficult to construct 
any bound state of fermions. 

Indeed, the results calculated by Bethe Ansatz method which is an exact method 
in two dimensional field theory model show that there is no boson state in the massless 
Thirring model.

In this respect, we cannot take the assumption of the existence of the NG scalar field 
for fermion field models from the physical point of view. 
Therefore, the Goldstone theorem does not hold for the case of fermion field theory, and 
the Nambu-Goldstone boson does not exist even after the chiral symmetry breaking of 
the NJL and Thirring models.

The absence of the NG boson is natural for the symmetry breaking of the two dimensional 
field theory since any massless boson should not exist. 
Therefore, the Thirring and the Gross-Neveu models have a chiral broken phase, and 
as the result, the fermion seems to acquire its mass. 

It is important to comment, here, on the condensate value with respect to the chiral 
symmetry breaking. The chiral symmetry breaking of the vacuum must be judged 
by the following equation,
\begin{equation}
 \hat{\cal Q}_5 |\Omega \rangle \not= 0
\label{intr_chiral_ch}
\end{equation}
where $\hat{\cal Q}_5$ denotes the chiral charge operator and the $|\Omega \rangle  $ 
is the symmetry broken vacuum state. 
Further, in this case, eq.(\ref{old_scalar_ev}) becomes
\[ \Phi_0 =  \langle \Omega| \bar \psi \psi |\Omega \rangle .\]
Since, in the spontaneous chiral symmetry breaking, 
the chiral charge must be conserved, it commutes with the Hamiltonian. 
In this case, the exact vacuum state must be the eigenstate of the Hamiltonian, 
and therefore the chiral charge has the same eigenstate as the vacuum state  
of the Hamiltonian. If we take this fact into account, then it is easy to prove 
that the fermion condensate of the exact vacuum state must vanish. But at the same time, 
we can show that if we operate the chiral charge $\hat{Q_5}$ onto the exact vacuum state, 
then we obtain eq.(\ref{intr_chiral_ch}), which means that the chiral symmetry is 
indeed broken. 
 
In this sense, the equation (\ref{old_spec_lim}) employed in the Goldstone theorem 
has lost its meaning. Instead, one should discuss the symmetry breaking in terms of 
the operator-induced  equation. If one takes the expectation value of 
the operator-induced equation, then one obtains identical equation of "$0=0$" due to 
the orthogonality condition between the vacuum state and other states induced 
by the operators, as we will show in the later section.

\section{New picture of symmetry breaking }
\label{new_sbp}
\subsection{General interpretation}

In quantum mechanics, the system has finite degree of freedom, and therefore 
one representation is equivalent to another representation whenever they link each other 
in the unitary transformation. Therefore, we can construct a physical spectrum 
(Fock space) based on one definite vacuum. On the other hand, it is well-known that, 
in quantum field theory which is the system of infinite degrees of freedom, 
there are different Fock spaces in one theory. 
This means that it is possible that there exist degenerate vacuums and these vacuums 
are diagonal to each other.  In this case, we can represent the system 
in terms of the creation and annihilation operator for each vacuum. 
However, the constructed Fock spaces are not unitary equivalent to each other. 
Therefore, we must take one Fock space vacuum for physical observation. 
Here, it is important to note that when one Fock space is preferred 
in the possible spaces, the others are forbidden and never realized in nature.

We are lead to an interesting consequence in the spontaneous symmetry breaking phenomena 
by the possibility of the existence for the degenerate vacuums. 
In this phenomena, there are symmetry unbroken vacuum $|0 \rangle$ as well as 
the symmetry broken (degenerate) vacuums $|\Omega, \alpha \rangle$ which satisfy 
eq.(\ref{sym_charge_judg}). Here, the label $\alpha$ in the broken vacuum is 
some distinguishable label of degeneracy.

\begin{figure}[h]
\begin{picture}(150,100)
\put(40,70){$|0 \rangle$}
\put(5,5){Unbroken vacuum}

\put(70,75){\vector(3,1){30}}
\put(70,73){\vector(1,0){30}}
\put(70,71){\vector(3,-1){30}}
\put(70,69){\vector(3,-2){30}}

\put(130,90){$|\Omega, \alpha_1 \rangle$}
\put(130,70){$|\Omega, \alpha_2 \rangle$}
\put(130,50){$|\Omega, \alpha_3 \rangle$}
\put(140,30){$\vdots$}
\put(110,5){broken vacuum}

\end{picture}

\caption{The physical vacuum of the symmetry broken phase becomes one of 
the degenerate vacuums}
\label{fig_symm_break}
\end{figure}
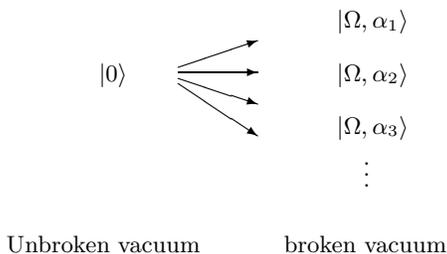

When the broken vacuum has a lower energy than unbroken vacuum, the physical vacuum 
becomes one of the degenerate symmetry broken vacuums, which is sketched 
in FIG.\ref{fig_symm_break}. In this case, the degenerate vacuum can be diagonal 
to each other,
\begin{equation}
\langle \Omega, \alpha|\,\Omega, \beta \rangle = \delta_{\alpha \beta}
\end{equation}
as well as the eigenstate of the Hamiltonian. Therefore, after one broken 
vacuum $|\Omega, \alpha_1 \rangle$ 
is selected by the symmetry breaking, the other vacuums  
$|\Omega, \alpha \rangle \,(\alpha \neq \alpha_1)$ 
are forbidden and never realized in nature except that some external disturbance 
are put into the system by hand. 
In this respect, the instanton picture in which the instanton can make transitions 
from vacuum to vacuum of the $\theta$ vacua in QCD is only a semi-classical picture. 

Now, we consider how the broken vacuum is parameterized. 
Let $\hat {\cal Q}_A$ be a charge (generator) of a considering symmetry. 
Let the system be described by the Hamiltonian $\hat H$ and the Hamiltonian be invariant 
under the symmetry. That is, 
\begin{equation}
\left[\hat {\cal Q}_A, \,\hat H \right]=0.
\label{new_qh_comm}
\end{equation}
Under the transformation which is generated by $\hat {\cal Q}_A$, 
the state of the system $|\Phi \rangle$ transforms to another state $|\Phi' \rangle$ as
\begin{equation}
|\Phi' \rangle = U(\hat {\cal Q}_A) |\Phi \rangle,
\end{equation}
where $U(\hat {\cal Q}_A)$ is a unitary operator of the symmetry. 
When the symmetry is spontaneously broken and the vacuum change from the symmetry 
unbroken (perturbative) vacuum $|0 \rangle$ to the symmetry broken vacuum 
$|\Omega \rangle$,  we can write
\begin{subequations}
\begin{align}
U(\hat {\cal Q}_A) |0 \rangle & =    |0 \rangle, \\ & \nonumber\\
U(\hat {\cal Q}_A) |\Omega \rangle & \neq |\Omega \rangle.
\end{align}
\end{subequations}
This means that
\begin{subequations}
\begin{align}
\hat {\cal Q}_A |0 \rangle & =    0, \\ & \nonumber\\
\hat {\cal Q}_A |\Omega \rangle & \neq 0.
\label{charge_vac1}
\end{align}
\end{subequations}
It is important to note that the Hamiltonian $\hat H$ and the charge $\hat {\cal Q}_A$ 
  \textit{always commute with each other} regardless the symmetry breaking. 
The conserved current and its associated charge are derived by the Noether theorem. 
How can we realize eq.(\ref{charge_vac1}) ? From the commutation relation 
eq.(\ref{new_qh_comm}) of  $\hat {\cal Q}_A$ and $\hat H$, one sees that 
the Hamiltonian and the charge are simultaneously diagonalized. 
Therefore, we can write the eigenvalue equations of the Hamiltonian $\hat H$ and 
the charge $\hat {\cal Q}_A$ as
\begin{subequations}
\begin{align}
\hat H|\Omega \rangle &=E_\Omega |\Omega \rangle, \\ & \nonumber \\
\hat {\cal Q}_A|\Omega \rangle &=Q_A |\Omega \rangle . 
\end{align}
\end{subequations}
That is, the vacuum is an eigenstate of the Hamiltonian as well as the charge operator 
$\hat {\cal Q}_A$. It means that the vacuum can be specified by the breaking charge as 
\begin{equation}
|\Omega \rangle = |E_\Omega, Q_A \rangle,
\end{equation}
where
\begin{equation}
\langle E_\Omega, Q_A'|\,E_\Omega, Q_A \rangle = \delta_{Q_A' \, Q_A}.
\end{equation}
Therefore, \textit{the broken vacuums are degenerate and are specified by the eigenvalue 
of the symmetry breaking charge}. Further, it should be noted that the eigenvalue 
of the continuous symmetry is not necessarily a continuous value. 
For example, the eigenvalue of the angular momentum in quantum mechanics has 
discrete values.

\subsection{Chiral symmetry breaking and Thirring model}

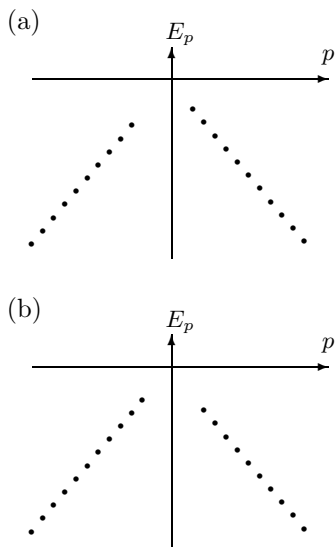
\begin{figure}[htb]
\setlength{\unitlength}{0.08pt}
\ifx\plotpoint\undefined\newsavebox{\plotpoint}\fi

\vspace{3mm}
\begin{minipage}{5cm}

\begin{picture}(1500,1100)(0,0)
\font\gnuplot=cmr10 at 10pt
\gnuplot
\sbox{\plotpoint}{\rule[-0.200pt]{0.800pt}{0.800pt}}%

\put(0,1100){(a)}
\put(779.0,12.0){\vector(0,1){1000}}
\put(779.0,82.0){\rule[-0.200pt]{0.400pt}{4.818pt}}

\put(120.0,860.0){\vector(1,0){1400}}
\put(891,728){\raisebox{-.8pt}{\makebox(0,0){\circle*{30}}}}
\put(944,666){\raisebox{-.8pt}{\makebox(0,0){\circle*{30}}}}
\put(997,604){\raisebox{-.8pt}{\makebox(0,0){\circle*{30}}}}
\put(1050,541){\raisebox{-.8pt}{\makebox(0,0){\circle*{30}}}}
\put(1102,479){\raisebox{-.8pt}{\makebox(0,0){\circle*{30}}}}
\put(1155,417){\raisebox{-.8pt}{\makebox(0,0){\circle*{30}}}}
\put(1208,355){\raisebox{-.8pt}{\makebox(0,0){\circle*{30}}}}
\put(1261,292){\raisebox{-.8pt}{\makebox(0,0){\circle*{30}}}}
\put(1313,230){\raisebox{-.8pt}{\makebox(0,0){\circle*{30}}}}
\put(1366,168){\raisebox{-.8pt}{\makebox(0,0){\circle*{30}}}}
\put(1419,106){\raisebox{-.8pt}{\makebox(0,0){\circle*{30}}}}
\put(604,653){\raisebox{-.8pt}{\makebox(0,0){\circle*{30}}}}
\put(551,590){\raisebox{-.8pt}{\makebox(0,0){\circle*{30}}}}
\put(498,528){\raisebox{-.8pt}{\makebox(0,0){\circle*{30}}}}
\put(445,466){\raisebox{-.8pt}{\makebox(0,0){\circle*{30}}}}
\put(393,404){\raisebox{-.8pt}{\makebox(0,0){\circle*{30}}}}
\put(340,341){\raisebox{-.8pt}{\makebox(0,0){\circle*{30}}}}
\put(287,279){\raisebox{-.8pt}{\makebox(0,0){\circle*{30}}}}
\put(234,217){\raisebox{-.8pt}{\makebox(0,0){\circle*{30}}}}
\put(182,155){\raisebox{-.8pt}{\makebox(0,0){\circle*{30}}}}
\put(129,92){\raisebox{-.8pt}{\makebox(0,0){\circle*{30}}}}
\put(1490,950){$p$}
\put(750,1050){$E_p$}
\end{picture}
\end{minipage}

\vspace{7mm}
\begin{minipage}{5cm}

\begin{picture}(1500,1100)(0,0)
\font\gnuplot=cmr10 at 10pt
\gnuplot
\sbox{\plotpoint}{\rule[-0.200pt]{0.800pt}{0.800pt}}%

\put(0,1100){(b)}
\put(779.0,12.0){\vector(0,1){1000}}
\put(779.0,82.0){\rule[-0.200pt]{0.400pt}{4.818pt}}

\put(120.0,860.0){\vector(1,0){1400}}
\put(944,666){\raisebox{-.8pt}{\makebox(0,0){\circle*{30}}}}
\put(997,604){\raisebox{-.8pt}{\makebox(0,0){\circle*{30}}}}
\put(1050,541){\raisebox{-.8pt}{\makebox(0,0){\circle*{30}}}}
\put(1102,479){\raisebox{-.8pt}{\makebox(0,0){\circle*{30}}}}
\put(1155,417){\raisebox{-.8pt}{\makebox(0,0){\circle*{30}}}}
\put(1208,355){\raisebox{-.8pt}{\makebox(0,0){\circle*{30}}}}
\put(1261,292){\raisebox{-.8pt}{\makebox(0,0){\circle*{30}}}}
\put(1313,230){\raisebox{-.8pt}{\makebox(0,0){\circle*{30}}}}
\put(1366,168){\raisebox{-.8pt}{\makebox(0,0){\circle*{30}}}}
\put(1419,106){\raisebox{-.8pt}{\makebox(0,0){\circle*{30}}}}

\put(653,716){\raisebox{-.8pt}{\makebox(0,0){\circle*{30}}}}
\put(604,653){\raisebox{-.8pt}{\makebox(0,0){\circle*{30}}}}
\put(551,590){\raisebox{-.8pt}{\makebox(0,0){\circle*{30}}}}
\put(498,528){\raisebox{-.8pt}{\makebox(0,0){\circle*{30}}}}
\put(445,466){\raisebox{-.8pt}{\makebox(0,0){\circle*{30}}}}
\put(393,404){\raisebox{-.8pt}{\makebox(0,0){\circle*{30}}}}
\put(340,341){\raisebox{-.8pt}{\makebox(0,0){\circle*{30}}}}
\put(287,279){\raisebox{-.8pt}{\makebox(0,0){\circle*{30}}}}
\put(234,217){\raisebox{-.8pt}{\makebox(0,0){\circle*{30}}}}
\put(182,155){\raisebox{-.8pt}{\makebox(0,0){\circle*{30}}}}
\put(129,92){\raisebox{-.8pt}{\makebox(0,0){\circle*{30}}}}
\put(1490,950){$p$}
\put(750,1050){$E_p$}
\end{picture}
\end{minipage}

\caption{The dispersion relation of the broken vacuum for the Thirring model. 
One has  $Q_5=1$ (a) and the other $Q_5=-1$ (b).
}
\label{fig_symm_broken_vac}
\end{figure}

As seen in previous subsection, the symmetry breaking phenomena can be described 
in terms of the vacuum's eigenvalue of the symmetry breaking charge. 
However, since the symmetry breaking phenomena should be a non-perturbative physics, 
we must find the vacuum  in a non-perturbative way. 
Unfortunately, it is difficult to find the non-perturbative vacuum of the quantum field 
theory even in an approximate scheme except for free fields. 
However, fortunately, the chiral invariant Thirring model which is a non-trivial field 
theory in two dimensions is exactly solved by the Bethe ansatz technique \cite{p5}. 
It is found that there are two types of vacuum. One is that the chiral symmetry is 
preserved and the other is that the chiral symmetry is not preserved. 
Further, the symmetry broken state has the lower energy than the symmetry preserved 
vacuum. Therefore, the physical vacuum of the Thirring model becomes a symmetry broken 
vacuum. We sketch the dispersion relations of the broken vacuum 
in FIG. \ref{fig_symm_broken_vac} whose analytic formulation \cite{p5} will be 
given in the next section.

Note that the chiral charge is the subtraction between the number of the left mover 
and the right mover fermions, 
\begin{equation}
Q_5 = N_L - N_R,
\end{equation}
where $N_{L,R}$ are the number of the left mover and the right mover, respectively. 
Indeed the number of the right mover and left mover fermions is different 
by one in the broken vacuum. Therefore, one corresponds to the eigenstate of 
the chiral charge $Q_5=1$ and the other corresponds to the charge $Q_5=-1$. 
Accordingly, the chiral symmetry of the Thirring model is spontaneously broken and 
the physical vacuum becomes one of the eigenstate of the chiral charge operator 
whose eigenvalue is $\pm 1$. After all, \textit{the vacuum of the Thirring model 
is classified by the chiral charge}.

\subsection{Coleman theorem and symmetry breaking phenomena in two dimensions}

The above results seem to be inconsistent with the Coleman theorem as well as Goldstone 
theorem. However, as we have learned above, there is an intrinsic problem 
in Goldstone theorem. The NG boson does not necessarily exist after symmetry breaking. 
Further,  it is important to note that the Coleman theorem cannot be verified  
since the vacuum expectation value of the scalar operator $\delta \phi$ becomes zero 
in two dimensions. 
Finally, we can conclude in usual sense that the chiral symmetry of the Thirring 
and Gross-Neveu models is spontaneously broken and fermion seems to acquire its mass.

However, there seems to be a problem since Coleman theorem indicates that
\begin{equation}
\langle \Omega | \bar \psi \psi | \Omega \rangle = 0 
\end{equation}
for the case of chiral symmetry breaking. The condensate value 
$\langle \Omega | \bar \psi \psi | \Omega \rangle$ 
has been believed to be an order parameter of the chiral symmetry breaking 
although there is no proof as to in which way the condensate value  can measure 
the chiral symmetry breaking. 

Surprisingly, we can find a good example for this case. This is the 
Thirring model. In the next section, we will show that the condensate value  of 
the Thirring model is  exactly zero even though the symmetry is spontaneously broken. 
Therefore, we conclude that the Coleman theorem in the Thirring model only states that 
the  condensate value vanishes in two dimensions.

\section{Chiral symmetry breaking and condensate}
\label{sec_chb_cond}

\subsection{General interpretation}
First, we review and reexamine the Goldstone theorem that has played a central role 
for understanding the symmetry breaking and its massless 
boson after the spontaneous chiral symmetry breaking \cite{q1,q2}. 
When the Lagrangian density has the chiral symmetry which can 
be represented by the unitary operator $U(\alpha)$, there is a conserved 
current associated with the symmetry 
\begin{equation}
 \partial_\mu j^\mu_5 =0 .
\end{equation}
In this case, there is a conserved chiral charge $ \hat{\cal Q}_5$ which is defined as 
\begin{equation}
 \hat{\cal Q}_5 = \int j^0_5 (x) d^3x .
\end{equation}
The Hamiltonian $H$ of this system is invariant under the unitary transformation $U(\alpha)$,
\begin{equation}
U(\alpha)H U(\alpha)^{-1} =H .
\end{equation}
Therefore, the chiral charge $\hat{\cal Q}_5$ commutes with the Hamiltonian $H$,  
\begin{equation}
 \hat{\cal Q}_5 H = H \hat{\cal Q}_5 .
\end{equation}
Now, the vacuum state can break the symmetry, and we define the symmetric vacuum 
$|0\rangle $ and symmetry broken vacuum $|\Omega \rangle $, respectively, which satisfy the following equations, 
\begin{subequations}
\begin{equation}
\hat{\cal Q}_5 |0\rangle =0,
\label{condensate_chiral_charge_unb}
\end{equation}
\begin{equation}
\hat{\cal Q}_5 |\Omega \rangle \not= 0 .
\end{equation}
\end{subequations}
Since the $\hat{\cal Q}_5$ commutes with the Hamiltonian, the  $\hat{\cal Q}_5$ has the same eigenstate 
as the Hamiltonian. If we define the symmetry broken vacuum state $|\Omega \rangle $ 
by the eigenstate of the Hamiltonian $H$ with its energy eigenvalue $E_\Omega $, 
then we can write 
\begin{equation}
H|\Omega \rangle =E_\Omega |\Omega \rangle .
\end{equation}
In this case, we can also write the eigenvalue equation for the  $\hat{\cal Q}_5$ 
\begin{equation}
\hat{\cal Q}_5|\Omega \rangle =Q_5 |\Omega \rangle 
\label{condensate_chiral_eigeneq}
\end{equation}
with its eigenvalue $Q_5$. 
These equations should hold for the exact eigenstates of the Hamiltonian. 

In the Goldstone theorem, one starts from the following commutation relation 
which is an identity equation, 
\begin{equation}
 \left[\hat{\cal Q}_5, \int  \bar \psi(x) \gamma_5 \psi(x)d^3x \right] = -2 \int  \bar \psi(x)  \psi(x)d^3x .
\label{condensate_chiral_id}
\end{equation}
Now, if we take the expectation value of the symmetry broken vacuum $|\Omega \rangle $ which 
is the eigenstate of the Hamiltonian as well as $\hat{\cal Q}_5$, 
then we obtain for the left hand side as
\begin{widetext}
\begin{equation}
 \langle \Omega| \left[\hat{\cal Q}_5, \int  \bar \psi(x) \gamma_5 \psi(x)d^3x \right]
|\Omega \rangle =   \langle \Omega| Q_5 \int  \bar \psi(x) \gamma_5 \psi(x)d^3x - 
\left( \int  \bar \psi(x) \gamma_5 \psi(x)d^3x \right) Q_5 
|\Omega \rangle =0
\end{equation}
\end{widetext}
with the help of eq.(\ref{condensate_chiral_eigeneq}). 
This means that the right hand side of eq.(\ref{condensate_chiral_id}) must vanish, that is, 
\begin{equation}
\langle \Omega| \int  \bar \psi(x)  \psi(x)d^3x |\Omega \rangle =0 .
\label{condensate_zero_eq}
\end{equation}
Therefore, the exact eigenstate has no condensate in the symmetry broken vacuum. 
What does this mean ?  The relation of eq.(\ref{condensate_chiral_id}) has repeatedly been used, 
and if there is a finite condensate, then the symmetry of the vacuum 
must be broken since the left hand side of eq.(\ref{condensate_chiral_id}) vanishes due to 
eq.(\ref{condensate_chiral_charge_unb}) 
for the symmetric vacuum state. However, the exact eigenstate shows that the condensate 
must vanish if the vacuum is the eigenstate of the Hamiltonian even for the symmetry 
broken vacuum state. 

The way out of this dilemma is simple. One should not take the expectation value 
of the vacuum state. Instead, the index of the symmetry breaking in connection 
with the condensate operator $\hat \Gamma$,
\begin{equation}
\hat \Gamma \equiv \int \!\!d^3x \,\bar \psi(x)  \psi(x),
\end{equation}
should be the following equation
\begin{equation}
\hat \Gamma |\Omega \rangle = |\Omega' \rangle +C_1 |\Omega \rangle
\label{condensate_op_st}
\end{equation}
where $|\Omega' \rangle $ denotes an operator-induced state which is orthogonal 
to the $|\Omega \rangle $. 
$C_1$ is related to the condensate value. For the exact eigenstate which breaks 
the chiral symmetry, we find 
$$ C_1 =0 . $$ 

In this case, the identity equation of (\ref{condensate_chiral_id}) can be applied to the state  
$|\Omega \rangle $ and we obtain
\begin{equation}
(\hat{\cal Q}_5-Q_5 ) \int  \bar \psi(x) \gamma_5 \psi(x) d^3x |\Omega \rangle = -2 |\Omega' \rangle
\label{condensate_chiral_id2}
\end{equation}
with the help of eq.(\ref{condensate_chiral_eigeneq}). Eq.(\ref{condensate_chiral_id2}) indeed holds true for
 the exact eigenstate. 
It is now clear that one should not take the expectation value of eq.(\ref{condensate_chiral_id}).

\subsection{ Exact solution of Thirring model}

There is one example which perfectly satisfies the above requirements of 
the spontaneous chiral symmetry breaking and zero fermion condensate. 
That is the Bethe ansatz vacuum of the massless Thirring model \cite{p5,non}. 
In the Bethe ansatz solution, the vacuum state 
is obtained exactly \cite{b5,q42,q43,q13,b133}, 
and this vacuum state breaks the chiral symmetry, contrary to Coleman's theorem \cite{q3}. 
However, it is shown that there is no bosonic state, and it has only a finite gap. 
Therefore, there is no contradiction with the fact that there should not exist 
any massless boson in two dimensions, which is the basis of Coleman's theorem. 

Now, we denote the right and left mover fermion creation operators 
by $a^{\dagger}_k$, $b^{\dagger}_k$, respectively, and 
thus the vacuum state $|\Omega \rangle $  can be written as
\begin{equation}
|\Omega \rangle = \prod_{k^r_i} a^{\dagger}_{k^r_i}\prod_{k^\ell_j} b^{\dagger}_{k^\ell_j} |0_v \rangle
\label{ex_thirr_vacu}
\end{equation}
where $|0_v \rangle $ denotes the null vacuum state with 
$$ a_{k^r_i}|0_v \rangle =0, \qquad b_{k^\ell_j}|0_v \rangle =0  . $$
The Bethe ansatz solution tells us that the momenta
$k^r_i$ and  $k^\ell_j$ should satisfy the periodic boundary condition (PBC) equations. 
These PBC equations are solved analytically and the momenta 
$k^r_i$ for right mover and  $k^\ell_j$ for left mover are given as \cite{p5}
\begin{subequations}
\begin{equation}
 k^r_i = {2\pi n_i \over L} - {2(N_{0}+1) \over L} \tan^{-1} \left({g\over 2}  \right)
\end{equation}
for $n_i=-1,-2,\cdots, -N_0$.
\begin{equation}
k^\ell_1={2N_0 \over L} \tan^{-1} \left({g\over 2}  \right)
\end{equation}
for $n_1=0$,
\begin{equation}
k^\ell_j = {2\pi n_j \over L} + {2N_{0} \over L} \tan^{-1} \left({g\over 2}  \right)
\end{equation}
\end{subequations}
for $n_j=1,2,\cdots, N_0$. 

Here, $g$, $L$ and $N_0$ denote the coupling constant, 
the box length and the particle number in the negative energy state, respectively. 
The cutoff $\Lambda$ is defined as
$$ \Lambda={2\pi N_0\over{L}}.  $$
We note that the negative energy $E_v$ of the vacuum state can be written as
$$ E_v=  k^r_i \ \ \  {\rm for}\ \  {\rm right \ \  mover}$$ 
$$ E_v= - k^\ell_j  \ \ \  {\rm for}\ \  {\rm left \ \  mover}. $$ 

Now, the condensate operator $ \hat \Gamma$ can be written as
\begin{equation}
\hat \Gamma = \int \bar \psi(x)  \psi(x) dx = \sum_n (b^\dagger_n a_{n}+a^\dagger_n b_{n} ) .
\end{equation}
Therefore, eq.(\ref{condensate_op_st}) becomes
\begin{widetext}
\begin{equation}
 \sum_n (b^\dagger_{n}a_n +a^\dagger_{n} b_n ) |\Omega \rangle = 
 \sum_n \left\{ \prod_{k^r_i,k^r_i \not= n} a^{\dagger}_{k^r_i}\prod_{k_j^\ell} 
b^{\dagger}_{k_j^\ell} b^{\dagger}_{n} |0_v \rangle +\prod_{k^r_i} a^{\dagger}_{k^r_i}
\prod_{k^\ell_j,k^\ell_j \not= n} b^{\dagger}_{k^\ell_j} a^{\dagger}_{n} |0_v \rangle \right\} . 
\label{ex_thirr_cond}
\end{equation}
\end{widetext}
Clearly, the right hand side of eq.(\ref{ex_thirr_cond}) is different from the vacuum state 
of the Bethe ansatz solution of eq.(\ref{ex_thirr_vacu}), and therefore denoting the right hand side 
of eq.(\ref{ex_thirr_cond}) by $|\Omega' \rangle $, we obtain
\begin{equation}
 \hat \Gamma |\Omega \rangle =
 \sum_n (b^\dagger_{n}a_n +a^\dagger_{n} b_n ) |\Omega \rangle = |\Omega' \rangle . 
\end{equation}
Obviously, the $C_1$ of eq.(\ref{condensate_op_st}) is zero in the massless Thirring model, and 
indeed this confirms eq.(\ref{condensate_zero_eq}). 

It is now clear and most important to note that one cannot learn the basic dynamics 
of the symmetry breaking phenomena from the identity equation. If one wishes to study 
the symmetry breaking physics in depth, then one has to solve the dynamics 
of the field theory model properly.

\subsection{ Bogoliubov vacuum of Thirring model}

Now, the massless Thirring model is also solved by the Bogoliubov transformation 
method \cite{q9,q91,q10}. In this case, the fermion condensate has a finite 
value as 
$$ \langle \Omega| \int \bar \psi(x)  \psi(x) dx |\Omega \rangle =
{\Lambda L\over{g\sinh \left({\pi\over g}\right)}} . $$
Further, the calculation shows that there is one massive boson. 
Comparing this result with the Bethe ansatz solution, we get to know 
where the Bogoliubov approximation goes wrong as far as the condensate 
evaluation is concerned. That is, the Bethe ansatz solutions show that 
the negative energy states of the left and right mover fermions are not mixed up 
with each other while the energy dispersions of  the two fermions 
in the Bogoliubov vacuum are connected to each other. 
This causes the finite fermion condensate in the Thirring model in the Bogoliubov 
transformed vacuum which is not the eigenstate of the Hamiltonian.

\section{Condensate and two dimensional gauge theory}
\label{condensate_gauge}
\subsection{ Schwinger model}

The Schwinger model is known to be solved exactly and can be converted 
to a free massive boson Hamiltonian. The mass of the Schwinger boson is given as 
\begin{equation}
 {\cal M}={g\over{\sqrt{\pi}}} 
\end{equation}
where $g$ denotes the gauge coupling constant. 

It also breaks the chiral symmetry. 
However, in this model, the chiral charge $\hat{Q_5}$ is not conserved due 
to the anomaly \cite{sch,m1,tom}. Therefore, the symmetry breaking is 
not spontaneous, and thus we cannot discuss the symmetry breaking in the same 
way as we discussed in section \ref{sec_chb_cond}. The basic point is of course that 
the chiral charge does not commute with the Hamiltonian. The Schwinger model 
has indeed a finite condensate value
\begin{equation}
 \langle \Omega| {1\over L}\int \bar \psi(x)  \psi(x) dx |\Omega \rangle =0.283 {g\over{\sqrt{\pi}}} .
\end{equation}
In this respect, the finite fermion condensate of the Schwinger vacuum is not 
inconsistent with the chiral symmetry breaking as discussed in section \ref{sec_chb_cond}. 
However, the condensate value is just the same as the one that is calculated 
by the vacuum state which is obtained from the Bogoliubov transformation. 
Since the Bogoliubov vacuum is not exact, it is not very clear whether 
the condensate vaule is exact or not. In any case, it should be carefully examined 
whether the fermion condensate of the vacuum state has any physical meaning or not.

\subsection{ QCD in two dimensions   }

Recently, careful studies of QCD$_2$ in two dimensions have been carried out by employing 
the Bogoliubov transformation method \cite{qq1}. The calculated results show that 
the vacuum has the finite fermion condensate which agrees very well with the calculation 
of the $1/N_c$ expansion method. This condensate value is given as
\begin{equation}
 \langle \Omega| {1\over L}\int \bar \psi(x)  \psi(x) dx |\Omega \rangle 
=-{N_c\over{\sqrt{12}}} \sqrt{N_cg^2\over{2\pi}} .
\end{equation}
The spectrum obtained in \cite{qq1} shows that there is one bound state (a massive boson), and 
the boson mass is expressed as
\begin{equation}
 {\cal M}_{N_c}={2\over 3}\sqrt{{N_cg^2\over{3\pi}}} 
\end{equation}
where the formula must be good for large $N_c$ region. 

From the argument in section \ref{sec_chb_cond}, the fermion condensate must vanish for QCD$_2$ 
if it is solved exactly. Up to now, the vacuum of QCD$_2$ has been solved 
only in  terms of the $1/N_c$ expansion approximation or the Bogoliubov transformation. 
It would be extremely interesting to find the exact vacuum state of QCD$_2$, 
and this is indeed a good problem to be solved in future.

\section{Conclusions}
\label{conclusion}
In the recent papers, we have shown that the Goldstone theorem in the fermion 
field theory models does not lead to the existence of the massless boson 
after the spontaneous chiral symmetry breaking \cite{q9,q91,qq1,p5,hist,nova}. 
There, we started from the finite fermion condensate value and showed that the identity 
equation of eq.(\ref{condensate_chiral_id}) cannot be used for proving the appearance of the massless boson. Also, it has been shown by numerical calculations that 
there is indeed no Goldstone boson in the fermion field theory models of the NJL, 
Thirring and QCD$_2$ after the spontaneous symmetry breaking takes place. 

Here, it is even more surprising that the exact eigenstate of the Hamiltonian 
leads to the conclusion that the fermion condensate must vanish 
for the symmetry broken vacuum state. Therefore, we cannot make use of 
the identity equation of the Goldstone theorem since the right hand side of the equation, 
that is, the fermion condensate vanishes for the exact symmetry broken vacuum. 

In this paper, we have shown that the Bethe ansatz solutions of the Thirring model 
present a perfect example that satisfies all the necessary and sufficient conditions 
in the spontaneous chiral symmetry breaking and its vacuum structure. The fermion condensate 
vanishes for the exact vacuum which breaks the chiral symmetry. The identity equation 
of the Goldstone theorem holds as the operator-induced equation, but has no meaning  
as the expectation value.  All the requirements of the symmetry breaking phenomena 
are realized in the Bethe ansatz vacuum of the Thirring model. 

Unfortunately, there is no exact solvable model in four dimensions. 
However, the NJL model is quite similar to the Thirring model 
apart from the renormalizability in the perturbative calculation,  
and  the solution of the NJL model must be quite similar to that 
of the Thirring model, concerning the zero condensate and a finite gap in the excitation 
spectrum. 

Further, we should comment on QCD in four dimensions since real nature must be described 
by QCD in four dimensions. However, there is no chiral symmetry in QCD in four dimensions 
due to the finite mass of quarks, and therefore there is no point to discuss 
the chiral symmetry breaking in QCD in four dimensions.

\vspace{0.5cm}

We thank Prof. K. Fujikawa for useful discussions and helpful comments. 

\vspace{1cm}

\end{document}